\documentclass[a4paper,11pt]{article}
\usepackage{pos}
\usepackage{booktabs}
\newcommand{\mrule}{\specialrule{.4pt}{.5pt}{1pt}}

\usepackage{enumitem}
\setlist{nosep} % or 

\usepackage{siunitx}

\title{A GeV to TeV view of shell-type SNRs}
 \ShortTitle{A GeV to TeV view of shell-type SNRs}
\manuallySeparateAuthors
\author*{Henrike Fleischhack}

\affiliation{Catholic University of America, Department of Physics\\
  620 Michigan Ave. N.E., Washington, DC 20064, USA}

\affiliation{NASA Goddard Space Flight Center,\\
8800 Greenbelt Rd, Greenbelt, MD 20771, USA}

\affiliation{Center for Research and Exploration in Space Science and Technology,\\ 8800 Greenbelt Rd, Greenbelt, MD 20771, USA}

\author{ on behalf of the Fermi Large Area Telescope Collaboration and}
\forColl{HAWC} % W/O "Collaboration"

\emailAdd{fleischhack@cua.edu}

\abstract{Shock acceleration by the shells of supernova remnants (SNRs) has been hypothesized to be the mechanism that produces the bulk of Galactic Cosmic Rays, possibly up to PeV energies. Some SNRs have been shown to accelerate cosmic rays to TeV energies and above. But which SNRs are indeed efficient accelerators of protons and nuclei? And what is the maximum energy up to which they can efficiently accelerate particles? Measurements of non-thermal emission, especially in the gamma-ray regime, are essential to answer these questions.

The High-Altitude Water Cherenkov (HAWC) observatory, surveying the northern TeV gamma-ray sky, is currently the most sensitive wide field-of-view survey instrument in the VHE (very-high-energy, >100 GeV) range and has recorded more than five years of data. The Large Area Telescope (LAT) onboard the Fermi satellite has been surveying the GeV gamma-ray sky for more than ten years. Combining measurements from both instruments allows the study of gamma-ray emission from SNRs over many orders of magnitude in energy. In this presentation, I will show measurements of VHE gamma-ray emission from Fermi-LAT-detected SNRs with the HAWC Observatory.
}

\FullConference{37$^{\rm{th}}$ International Cosmic Ray Conference (ICRC 2021)\\
		July 12th -- 23rd, 2021\\
		Online -- Berlin, Germany}

\begin{document}
\maketitle

\section{Introduction}

Shell-type supernova remnants (SNRs) have been considered as potential sources of Galactic cosmic rays due to two main arguments (see e.g. \cite{HEWITT2015674} and references therein). First, diffusive shock acceleration provides a mechanism for the expanding SNR shell to accelerate charged particles to relativistic speeds. Second, given the estimated rate of supernova explosions in the galaxy, the average energy released, and the predicted acceleration efficiencies of SNRs, these objects are expected to have sufficient power to sustain the measured intensity of Galactic cosmic rays.

In particular, young SNRs (hundreds of years old) are thought to be efficient accelerators of cosmic rays. Middle-aged SNRs (up to tens of thousands of years) can still be surrounded by a population of previously accelerated cosmic rays. However, it is unclear if SNRs do indeed have the ability to accelerate particles up to the so-called ``knee'', a break in the cosmic-ray energy spectrum around a few PeV, which is thought to indicate the maximum energy in the \emph{Galactic} component of cosmic rays. 

%Shock acceleration predicts approximately power-law energy spectra up to some maximum energy [needs to be expanded]

Any SNR that accelerates cosmic-ray protons to relativistic energies should also produce non-thermal gamma-ray emission due to interactions of these protons with the interstellar medium in/near the remnant. These interactions produce (among other things) neutral pions, which in turn decay into gamma rays. This emission mechanism is referred to as \emph{hadronic} emission. The gamma-ray energy spectrum is predicted to follow the underlying cosmic-ray energy spectrum. Any feature in the cosmic ray spectrum such as a break or cutoff should also be seen in gamma rays, at slightly lower energies. Thus, we can use the gamma-ray energy spectrum from GeVs to hundreds of TeV to investigate cosmic-ray acceleration up to PeV energies in SNRs. 

SNRs accelerating electrons to relativistic energies are also expected to produce non-thermal gamma-ray emission due to inverse Compton scattering of the CMB (or in some cases, ambient infrared or optical photon fields) and/or Bremsstrahlung processes in the presence of matter. These electrons are also expected to emit Synchrotron radiation at lower wavelengths (radio to X-ray, depending on the ambient magnetic fields). These emission mechanisms are referred to as \emph{leptonic} emission. In fact, most known SNR shells have been detected in radio surveys. 

HAWC is a large-field-of-view gamma-ray observatory sensitive in the energy range from hundreds of GeV to hundreds of TeV. HAWC has been surveying the northern TeV gamma-ray sky since 2015 and in its latest catalog reports 65 gamma-ray sources \cite{Albert_2020_3HWC}. 

The Large Area Telescope (LAT) aboard the \emph{Fermi} satellite has been surveying the gamma-ray sky in the energy range from tens of MeV to 2 TeV for more than ten years now. The 10-year \emph{Fermi}-LAT data release (4FGL-DR2, \cite{Abdollahi_2020}) contains 24 firmly \emph{identified} shell-type SNRs and an additional 19 sources that are less firmly \emph{associated} with SNRs, as well as 96 sources classified as ``spp'' type, indicating potential association with a PWN and/or SNR. At least three SNRs (IC 443, W44, and W51C) show a significant spectral feature, the ``pion bump'' at around 100 MeV, indicating that the gamma-ray emission is dominated by hadronic processes \citep{Ackermann_2013, Jogler_2016}. However, for many other SNRs, it is currently not clear if the gamma-ray emission is dominated by hadronic or leptonic processes. 

Several SNRs have also been detected at TeV energies. As of 2021/02/16, TeVCat\footnote{\url{http://tevcat.uchicago.edu/}}, a listing of TeV gamma-ray sources seen by HAWC and other ground-based gamma-ray observatories such as the imaging air-Cherenkov telescopes H.E.S.S., MAGIC, and VERITAS, lists 16 VHE gamma-ray sources associated with shell-type SNRs and 11 sources ascribed to the interaction of SNRs with nearby molecular clouds. However, many of these sources show spectral softening or a cutoff between GeV and TeV energies, indicating that those SNRs may \emph{not} be able to accelerate cosmic rays up to PeV energies. 

Here, we perform a search for TeV gamma-ray emission from GeV-emitting SNRs, focussing on SNRs that are not significantly detected by HAWC. Upper limits on the TeV gamma-ray flux will be determined. For SNRs where the upper limits are below the extrapolation from the GeV gamma-ray spectrum, we will have shown the presence of a break or cutoff in the spectrum, and can relate upper limits on the cutoff energy in the gamma-ray spectrum to the cutoff energy in the underlying proton or electron spectrum.

This contribution describes the source selection and analysis method; full results will be provided in a separate, peer-reviewed paper (currently under preparation).

\begin{figure}
\includegraphics[width=\textwidth]{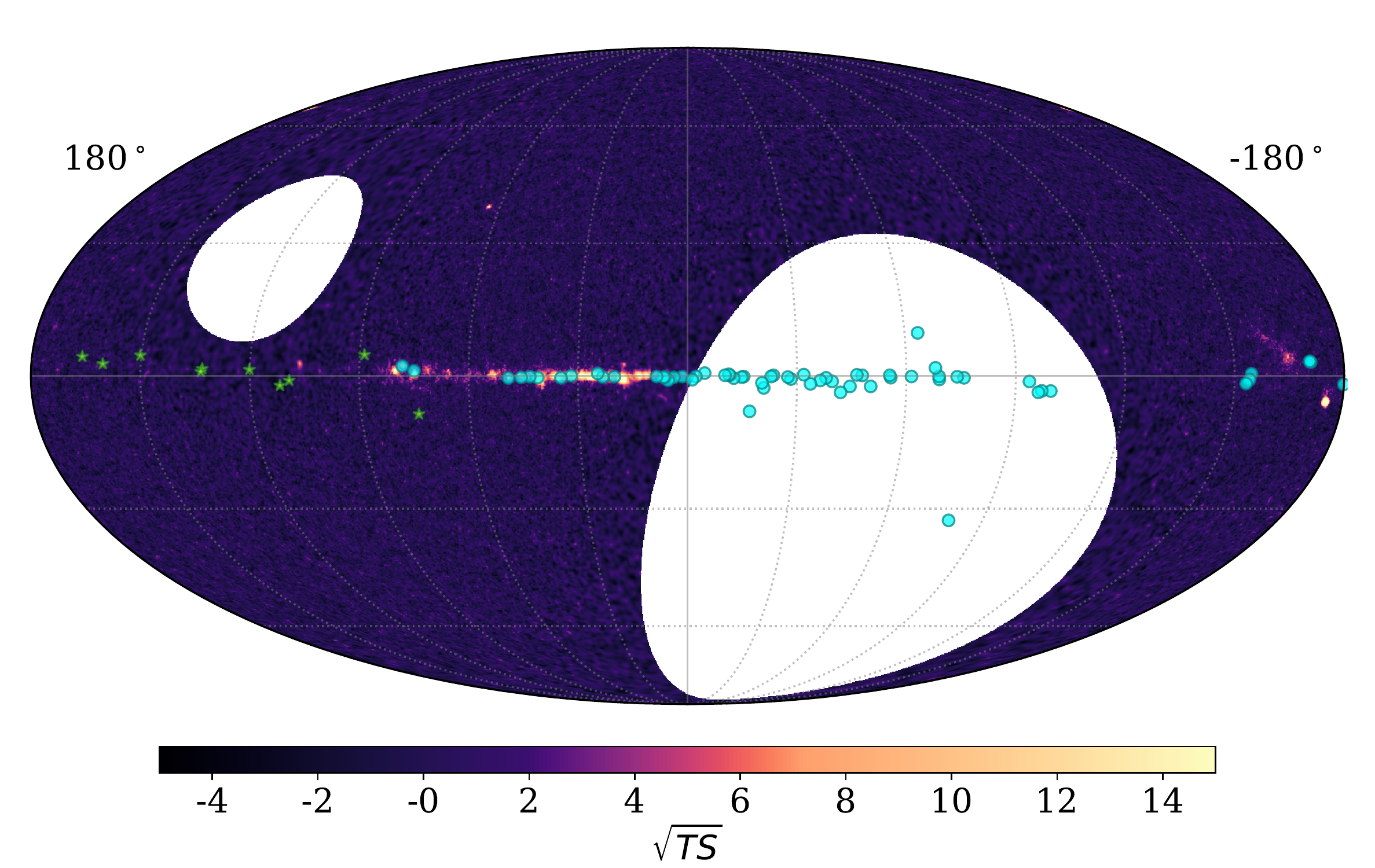}
\caption{HAWC's all-sky significance map for the point source search, based on 1523 days of observation, in Galactic coordinates \citep{Albert_2020_3HWC}. White areas of the sky are not observed by HAWC. Green stars mark the position of shell-type SNRs and SNR candidates from the LAT catalogs as described in Section \ref{sec:selection}. Aqua circles mark SNRs and SNR candidates that do not pass the selection criteria (outside of HAWC's field of view or overlapping with significant gamma-ray emission seen by HAWC).}
\label{fig:map}
\end{figure}

\section{Data and Analysis Methods}

\subsection{Source Selection}
\label{sec:selection}

GeV-detected shell-type supernova remnants (SNRs) and SNR candidates were selected from the following four catalogs released by the \emph{Fermi}-LAT collaboration:

\begin{itemize}
\item The LAT 10-year Source Catalog (4FGL-DR2, \cite{Abdollahi_2020,ballet2020fermi}),
\item The Third Fermi-LAT Catalog of High-Energy Sources (3FHL, \cite{Ajello_2017}),
\item Extended Sources in the Galactic Plane (FGES, \cite{Ackermann_2017}), and
\item The First LAT Supernova Remnant Catalog (1SC, \cite{Acero_2016}).
\end{itemize}

Sources were selected from these catalogs as follows:

\begin{itemize}
\item 1SC sources with the ``classification'' property given as ``classified'' (meaning, classified as SNRs).
\item 4FGL-DR2 and 3FHL sources with source type ``SNR'' (firmly identified as a shell-type SNR) or ``snr'' (less firmly associated with a shell-type SNR).
\item 3FHL sources associated with a 4FGL-DR2 SNR/snr source (via the ``ASSOC\_FHL'' property)
\item FGES sources not associated with a known PWN, given the associations listed in the FGES paper and the source types in the 3FHL, 4FGL-DR2, and TeVCat \footnote{http://tevcat.uchicago.edu}.
\item 3FHL and 4FGL-DR2 sources of type ``spp''\footnote{``Sources of unknown nature but overlapping with known SNRs or PWNe and thus candidates to these classes.''} within \ang{0.3} of a previously selected FGES or 1SC source.
\end{itemize}

Sources were required to have declinations (J2000) between \ang{-25} and \ang{65} to lie within HAWC's field of view.

Using the HAWC significance maps upon which the 3HWC catalog \citep{Albert_2020_3HWC} was based, SNRs were selected that did not overlap with regions where HAWC detects significant gamma-ray emission. 

%This includes SNRs in the inner Galaxy, IC 443, gamma Cygni, the three sources FGES~J0619.6+2229, 4FGL~J0639.4+0655e (Monoceros loop) and SNR205.5+00.5, which overlap/are near the TeV Halo candidate HAWC~J0635+070 (although likely not associated with it due to different morphologies). 
%FGES~J0537.6+2751/3FHL~J0537.6+2751e/SNR180.0-01.7/4FGL~J0540.3+2756e (all associated with Sim 147) were excluded as well due to their relatively large extent (\ang{1.5}) and proximity to the bright Crab nebula (\ang{5} distance to the center of Sim 147).

As all four LAT catalogs used different datasets and different energy ranges, a single SNR may be detected in only a subset of the catalogs. Additionally, SNRs detected in multiple catalogs may be found at slightly different locations or with different morphologies. Sources within \ang{0.3} of each other were considered to be in the same ``region'' if their extensions were also similar, which each region likely corresponding to the same physical source.

Table \ref{tab:sources} shows the ten selected SNRs and SNR candidates. Several are found in more than one catalog, leading to 21 sources to be studied. 

\begin{table}
\caption{SNRs and SNR candidates selected as described in Section \ref{sec:selection}. RA and Dec are the Right Ascension and Declination of the source center (J2000 epoch).  Ext is the exension (radius for sources modeled with a disk morphology, approximate radius or semi-major axis of a circle or ellipse covering the emission region for sources with irregular morphologies). Thin lines delineated between ``regions'' (e.g., the first three sources are all presumed to be associated with the same physical object).}
\label{tab:sources}
\begin{center}
\begin{small}
\begin{tabular}{llrrrlr}
\toprule
              Name &           Association &     RA [\si{\degree}]&   Dec [\si{\degree}] &  Ext. [\si{\degree}] &           Shape &   $\pm\sqrt{TS}$  \\
\midrule
3FHL J2051.0+3040e &     Cygnus Loop & 312.75 & 30.67 & 1.65 & CygnusLoop.fits &  0.60   \\
4FGL J2051.0+3049e &    Cygnus Loop & 312.75 & 30.83 & 1.65 & CygnusLoop.fits &  0.44     \\
     SNR074.0-08.5 &        & 312.77 & 30.90 & 1.74 &            Disk &  0.66       \\ \mrule
     SNR089.0+04.7 &           & 311.15 & 50.42 & 0.97 &            Disk & -0.07      \\
4FGL J2045.2+5026e &         HB 21 & 311.32 & 50.44 & 1.19 &            Disk & -1.34      \\ \mrule
     SNR109.1-01.0       &  & 345.41 & 58.83 & 0.00 &             & -0.41  \\
4FGL J2301.9+5855e &         CTB 109 & 345.49 & 58.92 & 0.25 &            Disk & -0.50\\
3FHL J2301.9+5855e &     CTB 109 & 345.49 & 58.92 & 0.25 &            Disk & -0.50 \\
 FGES J2302.0+5855 &     &  345.49 & 58.92 & 0.25 &            Disk & -0.50  \\ \mrule
     SNR111.7-02.1 &            & 350.85 & 58.83 & 0.00 &             &  0.46 \\
 4FGL J2323.4+5849 &      Cas A & 350.86 & 58.82 & 0.00 &                 & -0.45   \\
 3FHL J2323.4+5848 &       Cassiopeia A & 350.87 & 58.82 & 0.00 &                 &  0.44 \\ \mrule
 4FGL J0025.3+6408 &         Tycho &   6.34 & 64.15 & 0.00 &                 & -1.21  \\
 3FHL J0025.5+6407 &           Tycho &   6.38 & 64.13 & 0.00 &                 & -1.20  \\ \mrule
4FGL J0221.4+6241e &          HB 3 &  35.36 & 62.69 & 0.80 &            Disk & -0.01     \\ \mrule
4FGL J0222.4+6156e &       W 3 &  35.62 & 61.94 & 0.60 &         W3.fits & -0.52 \\ \mrule
4FGL J0427.2+5533e & SNR G150.3+04.5 &  66.82 & 55.55 & 1.51 &            Disk & -0.22   \\
3FHL J0427.2+5533e &      SNR G150.3+4.5 &  66.82 & 55.55 & 1.51 &            Disk &  1.10  \\
 FGES J0427.2+5533 &                &  66.82 & 55.55 & 1.52 &            Disk &  1.11 \\ \mrule
4FGL J0500.3+4639e &              HB 9 &  75.08 & 46.66 & 1.00 &        HB9.fits & -1.28    \\ \mrule
 4FGL J0526.7+4254  & SNR G166.0+04.3 &  81.69 & 42.92 & 0.00 &                 &  0.00    \\
\bottomrule
\end{tabular}
\end{small}
\end{center}
\end{table}

\subsection{HAWC Data analysis}

\subsubsection{Dataset and Analysis Software}
The analysis presented here is based on 1523 days of data recorded by HAWC. The same dataset was used for 3HWC, the third HAWC catalog. More details about the dataset can be found in \citep{Albert_2020_3HWC}.

HAWC-internal software was used for data reduction (calibration, event reconstruction, background rejection, and binning). The multi-mission maximum likelihood (3ML) framework \citep{Vianello:2015wwa} with the HAWC accelerated likelihood plugin (HAL) \citep{hal} was used to analyze the binned data. Each source was fit separately, with a region of interest (ROI) centered on the source and with a radius of \ang{5} for point sources, \ang{5} + source extension (see Table \ref{tab:sources}) for extended sources.

\subsubsection{Source Modeling}
Each source was modeled according to the position and morphology (disk or external template) given in the relevant LAT catalog. All parameters related to the morphology were kept fixed in the analysis.

For the spectrum, two separate analyses were carried out (both using the spatial model as described above):

\begin{enumerate}
\item The spectral shape was modeled according to the relevant LAT catalog, with free normalization. A normalization parameter $s$ was introduced which scales the flux normalization relative to the extrapolation from the LAT energies. 
\item  The spectral shape was modeled according to the relevant LAT catalog, multiplied by an exponential cutoff $\exp\left(- \frac{E}{E_{c}}\right)$ with free cutoff energy $E_c$. In this case, all other parameters including the normalization were kept fixed.
\end{enumerate}

A forward-fold likelihood fit (see \citep{Vianello:2015wwa}) was performed to find the best-fit values of the normalization constant $s$ or the cutoff energy $E_c$. For the first case, we also determined the test statistic $TS =-2 \ln\left( \frac{L_0}{L_s} \right)$, where $L_s$ is the value of the likelihood at the best-fit value of $s$ and $L_0$ is the value of the likelihood without the source in question (i.e., with $s=0$). $\pm \sqrt{TS}$ then corresponds to the detection significance of a given source, where we chose the negative sign if and only if $s<0$.

\begin{figure}
\begin{center}
\includegraphics[width=0.7\textwidth]{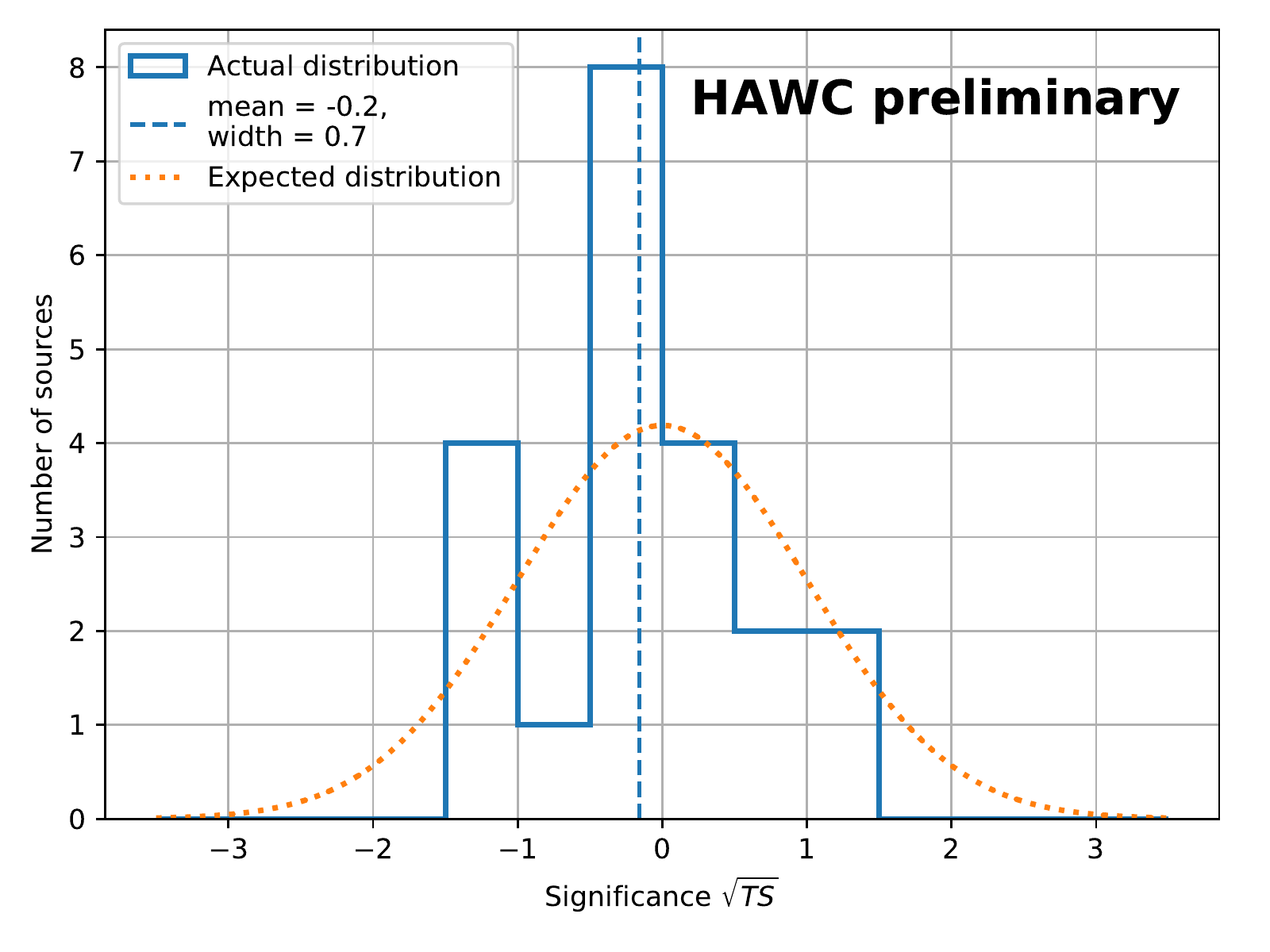}
\caption{Distribution of the detection significances from Table \ref{tab:sources}. The expected distribution is a Gaussian distribution with mean 0 and width 1. }
\label{fig:sig}
\end{center}
\end{figure}

\subsubsection{Upper Limit Determination}
As none of these sources were detected by HAWC, the most we can do is to set upper limits on the flux normalization. If the upper limit on $s$ is above $1$, that would mean that the non-detection of the given source is compatible with the extrapolation of the GeV spectrum to TeV energies (meaning, the source could be just too weak for HAWC to be able to see it) and we are not able to constrain its spectrum further. However, for sources with an upper limit on $s$ below one, we can conclude that the HAWC non-detection is inconsistent with the extrapolation of the GeV spectrum to TeV energies --- meaning that the spectrum must have a break or cutoff at TeV energies or above. We can try to set an upper limit on such a cutoff. 

In both cases (upper limit on the normalization scale factor $s$ and upper limit on the cutoff energy $E_c$), the limits were obtained from the likelihood profile. 90\% confidence level upper limits correspond to a 3.28 increase in the log-likelihood. 

\subsubsection{Systematic Uncertainties}
The systematic uncertainties due to the modeling of the HAWC detector were taken into account as described in \citep{Albert_2020_3HWC}, with the addition of the systematic uncertainty due to the source spectrum assumed in the preparation of the detector response file. For point sources, we also accounted for HAWC's systematic pointing uncertainty by letting the source positions float within the uncertainty and profiling over the source position when calculating the upper limit.

\section{Results}
Figure \ref{fig:sig} shows the significance distribution of \ref{fig:sig}. The distribution is compatible with a Gaussian distribution with mean 0 and width 1 as expected from the ``background-only'' case, indicating that there is no evidence for sub-threshold TeV emission in the selected sample of GeV-detected SNRs and SNR candidates.

\section{Outlook}

Full results including upper limits on the TeV emission from GeV detected SNRs will be presented in a dedicated publication (under preparation). 

%\clearpage

\bibliography{../bib}

\acknowledgments

We acknowledge the support from: the US National Science Foundation (NSF); the US Department of Energy Office of High-Energy Physics; the Laboratory Directed Research and Development (LDRD) program of Los Alamos National Laboratory; Consejo Nacional de Ciencia y Tecnolog\'ia (CONACyT), M\'exico, grants 271051, 232656, 260378, 179588, 254964, 258865, 243290, 132197, A1-S-46288, A1-S-22784, c\'atedras 873, 1563, 341, 323, Red HAWC, M\'exico; DGAPA-UNAM grants IG101320, IN111716-3, IN111419, IA102019, IN110621, IN110521; VIEP-BUAP; PIFI 2012, 2013, PROFOCIE 2014, 2015; the University of Wisconsin Alumni Research Foundation; the Institute of Geophysics, Planetary Physics, and Signatures at Los Alamos National Laboratory; Polish Science Centre grant, DEC-2017/27/B/ST9/02272; Coordinaci\'on de la Investigaci\'on Cient\'ifica de la Universidad Michoacana; Royal Society - Newton Advanced Fellowship 180385; Generalitat Valenciana, grant CIDEGENT/2018/034; Chulalongkorn University’s CUniverse (CUAASC) grant; Coordinaci\'on General Acad\'emica e Innovaci\'on (CGAI-UdeG), PRODEP-SEP UDG-CA-499; Institute of Cosmic Ray Research (ICRR), University of Tokyo, H.F. acknowledges support by NASA under award number 80GSFC21M0002. We also acknowledge the significant contributions over many years of Stefan Westerhoff, Gaurang Yodh and Arnulfo Zepeda Dominguez, all deceased members of the HAWC collaboration. Thanks to Scott Delay, Luciano D\'iaz and Eduardo Murrieta for technical support.

%% Full authors list (ONLY FOR COLLABORATIONS)
\clearpage
\section*{Full Authors List: \Coll\ Collaboration}
%
%\noindent \textbf{Note comment afterwards:} Collaborations have the possibility to provide an authors list in xml format which will be used while generating the DOI entries making the full authors list searchable in databases like Inspire HEP. For instructions please go to icrc2021.desy.de/proceedings or contact us under icrc2021proc@desy.de.\\
%

\scriptsize
\noindent
A.U. Abeysekara$^{48}$,
A. Albert$^{21}$,
R. Alfaro$^{14}$,
C. Alvarez$^{41}$,
J.D. Álvarez$^{40}$,
J.R. Angeles Camacho$^{14}$,
J.C. Arteaga-Velázquez$^{40}$,
K. P. Arunbabu$^{17}$,
D. Avila Rojas$^{14}$,
H.A. Ayala Solares$^{28}$
% ...

\noindent
% ...
$^{14}$Instituto de Física, Universidad Nacional Autónoma de México, Ciudad de México, México,
% ...
$^{17}$Instituto de Geofísica, Universidad Nacional Autónoma de México, Ciudad de México, México,
% ...
$^{21}$Physics Division, Los Alamos National Laboratory, Los Alamos, NM, USA,
% ...
$^{28}$Department of Physics, Pennsylvania State University, University Park, PA, USA,
% ...
$^{40}$Instituto de Física y Matemáticas, Universidad Michoacana de San Nicolás de Hidalgo, Morelia, Michoacán, México,
$^{41}$FCFM-MCTP, Universidad Autónoma de Chiapas, Tuxtla Gutiérrez, Chiapas, México,
% ...
$^{48}$Department of Physics and Astronomy, University of Utah, Salt Lake City, UT, USA,
% ...

\section*{No full author list for the Fermi-LAT collaboration is provided here in accordance with LAT collaboration policy for conference proceedings.}

\end{document}